\documentclass[oneside]{article}
\usepackage[T1]{fontenc}
\usepackage{authblk}
\usepackage{float}
\usepackage{amsmath}
\usepackage{graphicx}
\usepackage{xfrac}
\usepackage[english]{babel}
\usepackage{lmodern}
\usepackage[T1]{fontenc}
\usepackage[latin9]{inputenc}

\usepackage{csquotes}

\usepackage{mathtools}
\usepackage{graphicx}
\usepackage{esint}
\usepackage[unicode=true,
 bookmarks=false,
breaklinks=false,pdfborder={0 0 1},backref=false,colorlinks=false]
 {hyperref}
\usepackage[a4paper, total={210mm,297mm},margin=2.0cm]{geometry}

\title{The world beyond physics: how big is it?}

\author[1,2]{Sauro Succi \thanks{Electronic address: \texttt{sauro.succi@gmail.com}; Corresponding author}}

\affil[1]{Center for Life Nano-Neuro Science@La Sapienza, viale Regina Elena, 00161 Roma, Italy}
\affil[2]{Physics Department, Harvard University, Oxford Street 17, Cambridge, USA}

\date{\today}

\begin{document}

\maketitle

\begin{abstract}
We discuss the possibility that the complexity of biological systems 
may lie beyond the predictive capabilities of 
theoretical physics: in Stuart Kauffman's words, there is a 
World Beyond Physics (WBP). 
It is argued that, in view of modern developments of 
statistical mechanics,  the WBP is smaller than one might 
anticipate from the standpoint of fundamental physical theories.   
\end{abstract}


%


\section{Introduction}

Despite their vast differences in history, purpose, methods and 
procedures,  in the last decades physics and biology have taken 
significant steps towards each other. 
In particular, modern developments in non-equilibrium 
thermodynamics and statistical physics have 
shown increasing bearing on the understanding 
of the spacetime dynamics and evolution
of biological organisms \cite{EPS}.
The main theoretical thrust behind this endeavor is 
that the astounding progress in theoretical, experimental and 
computational methods has taken us to a position where we can explore many
instances of biological complexity on quantitative grounds.
Hence, in a optimistic vein, closing the gap between physics
and biology might be basically just a question of time.

On the other end of the spectrum, a few noted scholars 
maintain that the gap cannot be closed, by a matter of principle, because 
the "world is not a theorem" \cite{KAUF1,KAROL}. 

In his highly inspired and inspiring  book "A world beyond physics", 
Stuart Kauffman provides a series of passionate arguments on 
the reasons why biology cannot be reduced to physics.
In particular, he portrays physics as a discipline dealing 
with a machine-like world, as opposed
to the biosphere of living creatures which shape 
the laws they depend upon in ways which cannot
pre-stated and encapsulated in equations, rules or algorithms.

In this Perspective, I wish to argue that, while there is hardly 
any question that biology cannot 
be reduced to physics, (Phil Anderson said it best \cite{PWA}), 
the portrait of physics as a science devoted 
to a machine-looking world, does little justice to modern advances 
of theoretical physics, and most notably to the flourishing 
branch at the interface between physics, chemistry and biology, best known as 
soft matter and its theoretical underpinning,  that is
non-equilibrium statistical physics \cite{PCB}. 

In light of these developments, we argue that the WBP
appears significantly narrower than Kauffman's stimulating picture would have.

\section{The world beyond physics}

Kauffman argues that the complexity of the biosphere cannot be 
captured by physics because physics cannot anticipate
the emergence of structures whose very function is to promote 
and sustain their own existence (auto-poietic, Chapter 2,  Function of Function).
Of course,  such structures are contained within Newtonian phase space 
but they sit in an astronomically narrow corner which is not accessible
to Newtonian dynamics: the biological universe is not ergodic.
That's why the heart, and for that matter, any other biological 
organ, would live in a world beyond  
atoms and beyond physics,  Kauffmans' Kantian Whole. 
The argument is fair and square, save for a couple of key points; 
first,   there is (way) more to modern physics than Newtonian dynamics;
second, the ergodicity of Newtonian systems is still an open issue.
Before we spell the points out in some  detail, let us
first dig a bit deeper into the WBP.

\subsection{Kantian wholes}

In broad strokes, the hierarchy from Physics to Biology runs across 
the following multilevel sequence:
\begin{verbatim}
atoms -> molecules -> macromolecules ->  
CELLS -> tissues -> organs -> body
\end{verbatim}

This seven-level sequence encompasses ten decades in space, more than twice in time 
for a total of about hundred thousands Avogadro's atoms ($10^{28}$), running their show
over nearly hundred years on a time-clock ticking every femtosecond, thus 
covering about 24 decades in the process. 
This is the spacetime slice taken by a human being in his lifetime 
So, in purely reductionistic terms, the equation of the human body is basically 
a $10^{28}$ (quantum) many-body problem.
But is this purely reductionistic view operationally viable?  
The answer is a decided no.

A more sensible and productive way of handling the seven-level hierarchy is to ask how does
complexity grow as one walks across the various rungs of the ladder.
Is it monotonically growing or does it exhibit sharp growth or perhaps even
discontinuous jumps which would make Biology operationally 
and perhaps even ontologically disconnected from Physics?    
In principle, it is plausible to posit that the onset of cells
marks a boost in complexity,  the way as the formation of stars
marked a quantum leap of complexity in the history of our Universe.
This is another and perhaps more appropriate way of reformulating the basic
question of this Perspective: how big is the world beyond physics?
How far can physics reach into the Complexity of the above "Life chain"? 

Kauffman maintains that physics cannot keep up with the untamed rise of 
biological complexity: the heart exists as a Kantian Whole,  an entity that 
derives its very existence from the function it contributes to sustain: Life.
Even more radically: there's no formal system to describe biological evolution
because formal set theory cannot account for affordances \cite{GIB},  i.e.
the opportunities offered by the environment to the living creatures which
co-evolve with it.  
The issue implies a plunge in deep philosophical arguments far beyond this
authors's knowledge and ability. It may well be that the degree 
of contingency (affordances) we are used to in statistical physics, i.e.  initial and 
boundary conditions, various forms of stochasticity, fall short of accounting 
for the complexity of the different functions 
that a given organism can deliver. 
Yet, this does not prove that physical situations dominated 
by contingencies are outside the realm of formal mathematization.
Differently restated, the existence of the heart is obviously
compatible with the zillions of atoms that compose it, but
the question is: can the equations/algorithms that describe 
atoms predict its emergence?
To begin with, one should recognize that 
Newton's description is not adequate to answer this question, because its phase 
space is far too vast for systematic quantitative explorations.
Differently restated,  it is not known whether Newtonian trajectories
can ever reach those portions of phase spaces subject to the
constraints which control the emergence  of the heart 
from the world of atoms.
A similar argument,  in a much stronger form , applies to the many-body
Schroedinger wavefunction, which Walter Kohn defined
"not a scientifically legitimate concept" beyond some 
thousands particles, on account of its uncomputability \cite{KOHN}.
\begin{figure}
\includegraphics[scale=0.5]{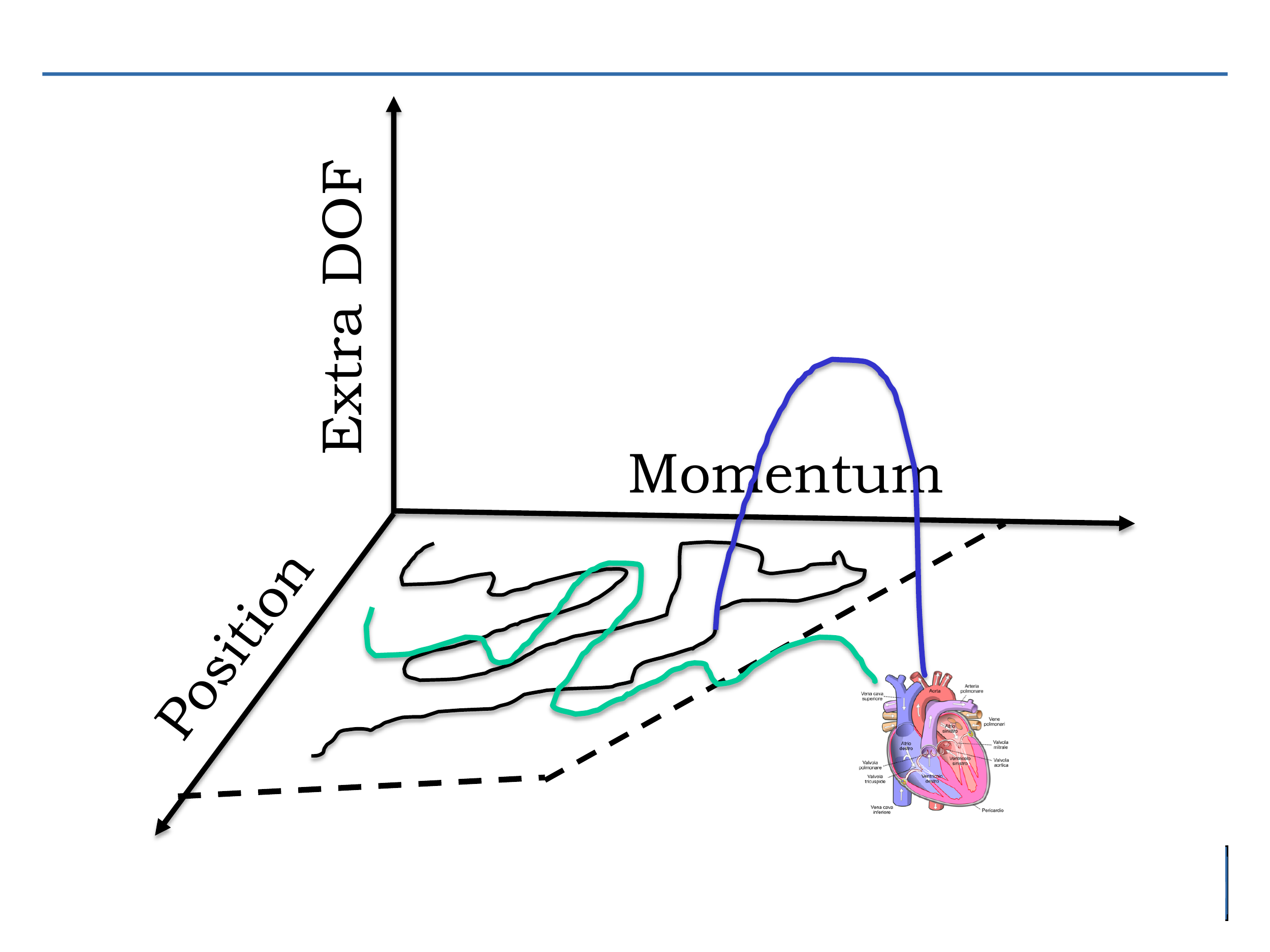}
\caption{Weak and Strong emergence.  
According to the WBP idea, the "heart corner", albeit part of Newtonian
phase-space,  cannot be reached by any Newtonian trajectory, no matter 
how long in time (black thin curve), because the biological universe is non ergodic.
According to the weak-emergence scenario discussed in this paper, Newtonian degrees
of freedom "coalesce" to form coarse-grained variables undergoing collective motion.
Such collective motion permits to explore phase-space much faster than Newtonian
mechanics, but since it is realizable, by definition, it cannot reach any 
region of phase-space which would not be accessible to Newtonian trajectories as well.
According to the strong-emergence picture, the heart can only be reached by taking a jump
along extra-dimensions beyond Newtonian phase space, described by novel degrees of freedom
which arise "on the fly" and cannot be prestated (horse-shoe segment in blue).
}
\end{figure}

But other descriptions may offer better insights.
For instance,  a strong advocate of a rule-driven universe 
is Stephen Wolfram \cite{WOLF},  based 
the "Complex from Simple" (CfS) paradigm:  simple rules,  indefinitely
iterated in time,  give rise to fairly complex behavior.
Cellular Automata (CA) algorithms stand as a poster child for this approach.
CA are powered by the notion of Universality,  namely the property by which
inessential details are washed out in the process of iterating the rules, while
they important ones are reinforced instead.
However,  with a few noted exceptions, CA fall short of providing a quantitative
description of real-life complexity.  In fact, it is quite possible 
that beyond a given level of complexity,  Nature may rather obey 
a more realistic "Complex from Complex" (CfC) paradigm.
Some authors note that, unlike physical systems, even the ones most credited 
as paradigms of  complexity,  such as spin-glasses \cite{SPIN}, complex 
(adaptive) systems co-evolve with their governing rules \cite{THU,YOLO,GROS}.
The point is interesting,  but would take us astray.  
For the sake of concreteness, we shall focus instead on just two 
central ideas which do not necessitate co-evolving rules: 
{\it coarse-graining} and {\it structure formation}.
Before plunging into these matters,  let us discuss the important 
notions of weak and strong emergence. 

\section{Weak and Strong Emergence}

The coarse-graining (CG) procedure to be described in the
next sections consists in formulating effective equations for collective variables 
resulting from the aggregation of many microscopic degrees of freedom.

CG is commonly pursued under the basic constraint of 
{\it Realizability}, i.e.  the collective dynamics must be 
compatible with  the underlying micro-dynamics and
lack of realizability is  regarded as a show-stopper for any CG model. 
This corresponds to the so-called "weak emergence", meaning by this that the
emergent interactions are still grounded in the underlying microscopic picture and cannot
produce any solution that could not be obtained,  {\it in principle},  by running the microscopic dynamics.

This contrasts with the so called "strong emergence", whereby 
one postulates the rise of new "unrealizable" interactions 
that {\it cannot} be traced back to the underlying microscopic ones.
According to the strong-emergence picture,  even if we could 
solve Newton's equations for an arbitrary
number of particles with arbitrary precision and indefinitely long in time
(a conceptual as well as practical chimera), we would not 
been able to tell why X fell in love with Y instead of Z.  
This requires new functional degrees of freedom which pop out "on the fly" during
the evolution of the system and are not contained in Newton's phase 
space, no matter how large.
For such systems, phase-space itself becomes a dynamic and growing  structure 
which co-evolves with its inhabitants \cite{TAP}.
This is an overly fascinating idea and possibly even a right one.
But complex systems rarely abide by the Keats 
rule "Beauty is Truth and Truth is Beauty" and  
in this Perspective we shall indeed argue in favor of the
more traditional weak-emergence picture
because of its logical economy: Occam instead of Keats.

Finally, let me emphasize that the strong-emergence portrayed above is 
stronger than the one discussed for instance in \cite{ELLIS},  in which
it is assumed that the collective degrees of freedom react back on the
microscopic ones (down-ward causation) but are still microscopically 
realizable. The coarse-graining picture described in the sequel is 
compatible with such "weak" form of strong emergence,  as detailed 
in the closing section of the manuscript.

\section{Coarse-graining}

Coarse-graining (CG) is the act of turning from a microscopic to a 
higher level of description, in our case a mesoscopic or macroscopic one.
By "meso" we mean an intermediate level between
the microworld of atoms and the macroworld of (classical) 
continuum fields.  Somehow, between  discrete molecules and  fluid density, pressure,
temperature and so on.  A world where probability takes central stage.
The genuinely new interactions which govern the CG procedure are often called 
{\it emergent}, since they literally emerge from the underlying
microscopic interactions by some form of statistical averaging.
Although they do not enjoy the same status of the fundamental
three,  electro-weak, strong and gravity,   they are much more 
effective in exploring the complex interface between physics 
and biology \cite{PWA, PCB}.    

The key is that once one shifts focus from individual 
to collective motion, which is precisely what 
statistical physics does,  genuinely new interactions arise, which are
well positioned to narrow down the gap between physics and biology
because they can converge to the regions of phase-space relevant to biological
evolution much faster than Newtonian dynamics could ever possibly do.
The claim here is that no co-evolving landscape is needed.

After such general premise, let us illustrate CG  in some more detail. 
The state of a classical (non-quantum) many-body system consisting of $n$ 
individual units (atoms for convenience) is described by a set of 
$n$ Newtonian equations of motion:

\begin{eqnarray}
\label{NEWTON}
\frac{dr_i}{dt} = v_i\\
m_i \frac{dv_i}{dt} = \sum_{j>i}  f(r_i,r_j),\;i=1,n
\end{eqnarray}

where $n$ is a large number, say of the order of the Avogadro number, $N_{av} \sim 10^{23}$. 
The key ingredient are the inter-atomic two-body (for simplicity) 
forces  $f(r_i,r_j)$ which,  albeit known from the fundamental
point of view (say electrostatics or gravity), give rise to 
an enormous variety of dynamical behaviours
just due to the fact that they are {\it many} and {\it nonlinear}. 
A glass of water contains an Avogadro numbers of molecules... 

Most importantly,  the rich variety of solutions depends critically 
on the boundary and initial conditions, a crucial
issue to which we shall return shortly. 
Since the many-body Newton's equations are far too many 
to solve,  one naturally seeks for cogent simplifications
which would not only retain the essential physics, but actually 
{\it distill} it out of the multitude of the possible
microsolutions,  thus cutting away the unvisited regions 
of phase-space (non-ergodicity) but also visit the accessible ones which
would remain unexplored by the microscopic dynamics for want of time.

A most drastic and yet highly informative mesoscale 
approximation of Newtonian mechanics is the so-called 
Langevin equation, in which the 
ensemble of $n$ Newtonian atoms is lumped 
into just one single "meso-particle",  
representing the average particle interacting with all other molecules 
through effective one-body forces .
In general, each meso-particle represents a collection of $\mathcal{B}$ 
molecules where the "blocking factor" $\mathcal{B}$ provides 
a quantitative measure of the degree of coarse-graining.
In this many-body mesoscale representation,  each mesoparticle 
obeys the following set  of generalized Langevin equations,  known 
as Dissipative Particle Dynamics (DPD) \cite{DPD1,DPD2}:

\begin{eqnarray}
\label{DPD}
\frac{dR_I}{dt} = V_I \\
M_I \frac {dV_I}{dt} = \sum_{J>I}^N (F_{IJ}^C +F_{IJ}^D+F_{IJ}^R) \;\;\;I,J=1,N=n/\mathcal{B}
\end{eqnarray}

In the above $R_I$ is the position of the $I$-th mesoparticle of mass 
$M_I=\sum_{i \in I}  m_i$ and $V_I$ is its velocity.
The mesoscale force consists of three components, labelled 
by $C$, $D$ and $R$ for Conservative, Dissipative 
and Random, respectively,  acting on the $I$-the mesoparticle  
as a result of the direct interaction with the $J$-th one.
\begin{figure}
\includegraphics[scale=0.5]{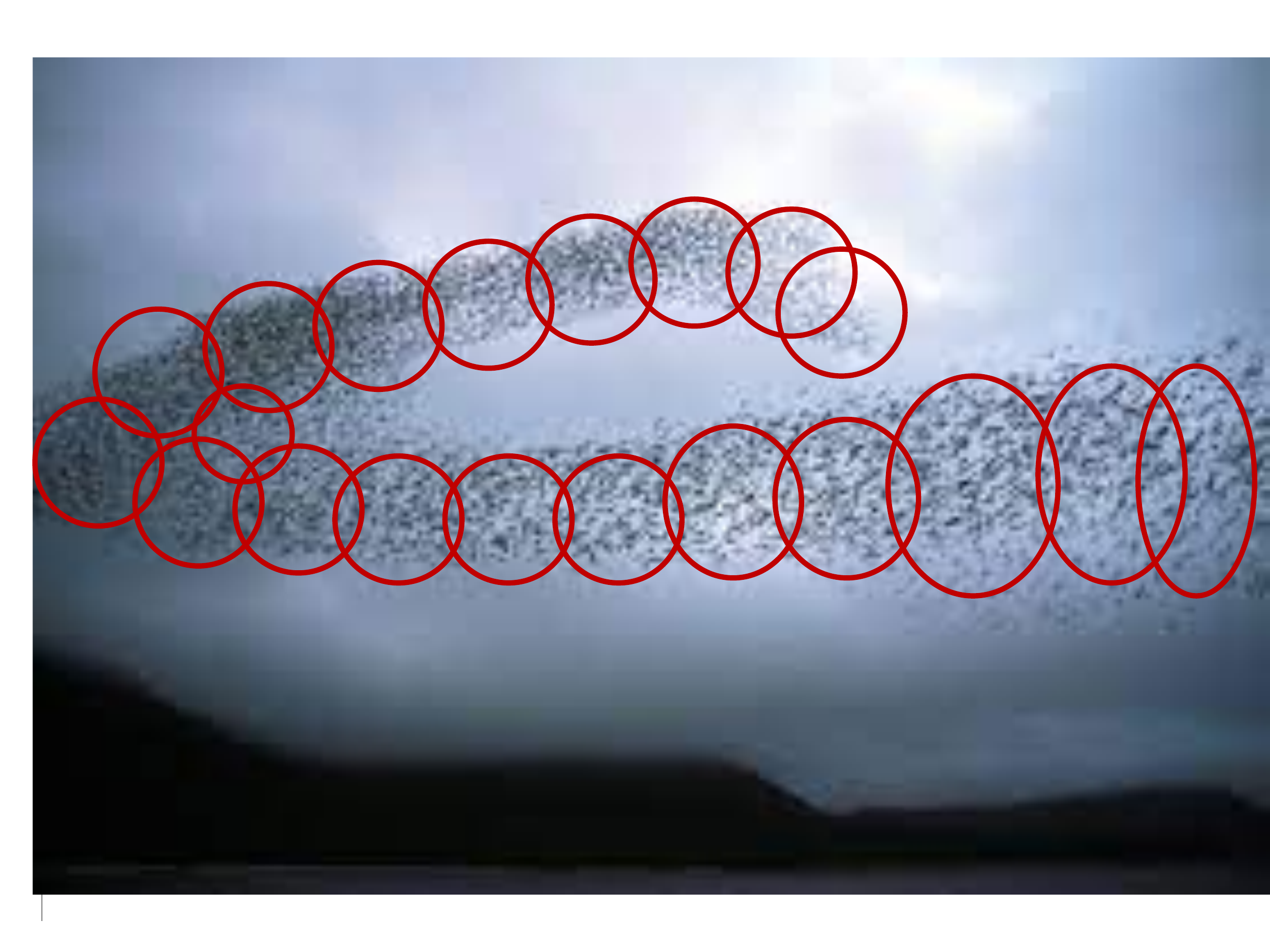}
\caption{The notion of collective variable (mesoparticles).
A swarm of thousands birds displays collective motion which takes birds to regions
of space that would remain inaccessible to each bird in isolation (useful to avoid predatory attacks).
This collective motion is best described in terms of a few "mesoparticles" (red circles), each of which
represents the dynamics of a large group of "correlated birds".  
}
\end{figure}

By its very nature,  compression of information brings to light
new interactions which would remain hidden in the forest of Newtonian details.
Incidentally,  nonlinearity is key since coarse-graining does not commute 
with nonlinearity, and the new interactions arise precisely from this lack
of commutativity.  A few comments are in order.
First, we note that the DPD equations still bear a strong resemblance 
to Newton equations, yet,  with the following {\it three} crucial twists:

1) {\it Conservative Forces}: 
The mesoscale force 
$$F^C_{IJ} \equiv F^C(R_I,R_J)$$  is generally different from its 
micro-counterpart $f(r_i,r_j)$ and usually more complicated.   
In the most fortunate instances, such difference boils down to a rescaling
of the interaction parameters (Renormalization), in which case the 
conservative force is called "minimal".  It should be borne in mind that 
Renormalization is a fortunate circumstance and by no means the rule. 
Hence,  the specific form of $F^C$ is mostly based on informed heuristics. 
To be noted that while the mass of the mesoparticles scales like $\mathcal{B}$,
the meso-forces scale at most like $\mathcal{B}^{2/3}$ because 
the internal forces within the particle cancel out by Newton's third law.
As a result, the acceleration scales like $\mathcal{B}^{-1/3}$, meaning that a given
distance can be covered by a larger timestep, scaling like $\mathcal{B}^{1/6}$.
This is a weak dependence, but with $\mathcal{B} \sim 10^9 \div 10^{12}$, it
implies a nearly two-order boost in the timespan of the trajectory at a given number of time steps.
Two orders of  magnitude are significant,  but astronomically short of accounting for the prodigious 
speedups displayed by fundamental biological processes,  such as protein folding.
To this purpose,  the mesoscale forces ought to display
highly non-trivial remappings of Newtonian forces,  such as the
emergence of free-energy funnels  \cite{WOLYN,SS22}, leading
to an exponential acceleration of the CG dynamics, which is
how biologically relevant regions of phase space may become accessible.

2) {\it Dissipative Forces}: The second term on the RHS is brand new,  
as it reflects a mechanism,  dissipation,  with no counterpart 
in the Newtonian world (non-minimal coupling).  
Dissipation is the result of the interaction of the 
mesoparticles with their environment (reservoir),  to which they 
irreversibly loose their momentum and energy. 
A popular choice is the simple linear expression:
$$
F_{IJ}^D = -\gamma  (V_I - V_J) e_{IJ}
$$
where $e_{IJ}$ denotes the unit vector along the direction joining 
particle $I$ to particle $J$. The role of dissipative forces is again crucial, for they
assure the convergence to the proper attractors.

3) {\it Fluctuating Forces}: The third term is also non-minimal,  as it reflects another 
mechanism which does not exist at the microscopic level: statistical noise,  due again 
to the erratic interaction of the mesoparticle with its fluctuating environment:
$$
F_{IJ}^R = \sigma  \xi_{IJ} e_{IJ}
$$
where $\xi_{IJ}$ is a random number in $[0,1]$ and $\sigma$ measures 
the strength of the fluctuations, i.e. the temperature of the system.
Note that Fluctuation and Dissipation are two faces of the same
medal,  linked by the so-called Fluctuation-Dissipation theorem, which
fixes $\gamma$ as a function of $\sigma$ and viceversa.
Random forces are essential to escape local minima in the (free) energy landscape.


Both dissipative and fluctuating forces may carry non-linear and non-local
dependencies in space and time (non-Markovian processes), because
the motion of the particle affects the environment, so that the next particle 
may find a different one  as compared to its forerunner. 
Since the environment is much larger  than the
particle,  it is expected to regain its equilibrium on a much shorter  timescale than 
the one of particle motion.   Whenever such assumption fails, 
memory effects (in space and time) take stage. 
That's how long-range correlations emerge from local microscopic interactions,  a
profound hallmark of complexity. 

Emergent interactions are no less 
fundamental than the microscale interactions they derive from. 
In fact, with reference to their own level of description, they are
actually {\it more} fundamental.
The point where they "lose" fundamental status is different, and 
namely the fact that while Newton's equations are self-consistent, 
$n$ equations for the $n$  degrees freedom, at the mesoscale
level self-consistency is lost and must be forced 
back in by a procedure which is bound to loose information.
A process known as {\it closure}.

In other words, the act of formally deriving the equation of 
motion for the mesoscale coordinates $R_I$ and $V_I$, inevitably 
generates additional variables (correlations), which in turn obey 
additional equations, along an open hierarchy which 
only closes once $N=n$ collective variables are included. 
At this point coarse-graining reduces to a "mere" change of 
coordinates,  which may be useful on its own,  but defeats the very spirit 
of the ordeal, namely cut down the  amount of information to be processed 
by retaining what is essential and discarding the unessential 
(the usual "baby in the tub" problem).
There are no exact procedures to accomplish the "perfect" CG,   except 
for very special instances,  which means that CG invariably comes
at the price of some irreversible loss of information.
This is precisely where CG picks up its "artistic" component: maximize
the return on investment by relinquishing unnecessary information 
without mangling the essential physics.
This is a powerful strategy for Nature to search the tiny
corners of phase-space "where things work" (the heart).
It is often emphasized that living systems are powerful
information processors, less noted perhaps that they are
also and perhaps primarily, fairly efficient CG-machines \cite{SS22}.

As to the ergodicity of Newtonian trajectories, rigorous results are exceptionally hard to obtain \cite{SINAI,GALLA} and
computers can help,  but only to a point, because of the infamous 
time-gap problem.  Tracing months-long trajectories (the time it takes for the heart to form) by 
ticking at a femtosecond scale,  means simulating of the order of $10^{21}$ steps.
Not only is this far beyond the capabilities of any foreseeable computer.
Some authors maintain that even 
if such computer were with us,  they would not help much because of 
dynamic instabilities as combined with round-off errors due
to the floating point representation of real numbers \cite{BOG}.
One may observe that the above hindrances have not  prevented the 
correct calculation of transport coefficients, nor did they hamper
the predictions of climate models, as recognized by the 2021 Nobel Prize in Physics,
but the point remains an interesting and important one.

The key issue though, is the capability of correlations to 
fast-drive the systems towards the "hidden" regions of phase space where biological
functions can flourish,  while ignoring the overwhelming majority of phase space
in which such functions stand no chance.


\section{Structure formation}

The ability of growing coherent structures by exchanging mass, energy and information
with the surrounding environment is one of the main hallmarks of living systems.
The standard mathematical paradigm for structure formation is the reaction-diffusion 
model pioneered by the epoch-making Turing's paper \cite{TUR} back in the 50's.
Mathematically this is described by a set of coupled reaction-diffusion 
equations of the form:
\begin{equation}
\label{DR}
\partial_t c_s = D_{ss'} \Delta c_{s'} + \mathcal{R}_s(c) 
\end{equation}
where $c_s$ is the concentration of species $s$, $D_{ss'}$ the corresponding
diffusion matrix and $\mathcal{R}_s$ is the rate of 
production/consumption of species $s$
due to chemical reaction with the other species.  
Details change from system to system and they matter a lot, but the basic mechanism 
is fairly general; chemistry is local and nonlinear, while diffusion is linear and non-local.  
Taken together, they manage to transfer energy, mass and momentum across 
scales thus giving  rise to a rich variety of non-equilibrium pattern formation phenomena.
In particular, chemistry selects the critical size above which structures 
can grow and survive against dissipation, laying the foundations for morphogenesis.

\subsection{Dissipative Structures and Beyond}

The key insight is that the coupling between non-linear local 
chemistry and linear non-local diffusion, gives rise to organized patterns in spacetime
which owe their stability (order) to the entropy they manage to dump to the 
environment \cite{WHAT}.
To quote Prigogine, "entropy is the price of structure" \cite{PRIGO1,PRIGO2}.

This insight extends to a broad variety of non-linear systems far from equilibrium,  a
paradigmatic example in point being Rayleigh-Benard cells, namely 
the coherent rolls which develop upon heating from below a fluid 
confined between two plates kept at constant temperature,
once a critical temperature difference 
between the lower (hot) and upper (cold) plates is exceeded.

Prigogine's passionate attempts to elevate dissipative structures to
the status of a paradigm of living systems,  didn't meet with much consensus among his peers.  
One of the reasons, as observed  by PW Anderson,  was the lack of a theory of 
dissipative structures far from equilibrium.
Indeed, Prigogine's principle of maximum entropy production, builds heavily on
Onsager's reciprocity,  which only holds near equilibrium.
Modern developments in non-equilibrium statmech, particularly the fluctuation theorems
by Crook, Jarzsynski and Gallavotti-Cohen \cite{CROOK,JAR,GALCO} considerably extend 
the scope for quantitative analysis of systems far from equilibrium.  
For instance,  J. England recently suggested "dissipative adaption"
as a general paradigm by which living systems adapt to time-varying external drives
by maximizing dissipation and quotes a number of self-assembling phenomena as
a practical manifestation of such "principle" \cite{JENG13,JENG15}.
England's grand-picture is very elegant but again,  whether such elegance 
captures the actual behavior of real life living systems remains open for grabs.

Interestingly,  Kauffman notes that one reason why DS fail to 
attain the level of complexity of living organisms is the fact that biological 
systems  spend useful work (free energy) to build up their 
own boundary conditions, the cell being a prime example in  point. 

By contrast,  dissipative structures are subject to fixed boundary 
conditions, hence they can only generate more structures of different 
sizes and energy,  but no qualitatively new structures at a higher level
of the complexity ladder. This is a truly insightful remark which, in my view, could 
stimulate genuinely new work in modern statmech.
This takes us to the next and final item: soft flowing reactive matter.

\section{Soft Flowing Reactive Matter}

Systems that spend work to build  their own boundary conditions and 
co-evolve jointly together are a commonplace in soft matter. 
Consider for instance the formation of supramolecular structures, such as 
membranes, micelles and others,  from the mesoscale dynamics of the system.
In these models, one specifies suitable effective interactions which drive the
formation of such supramolecular structures due to the combined effect of many
nonlinear physical mechanisms,  like surface tension(capillary forces), near-contact 
interactions of various sorts \cite{THERMO,LB18,AM19} and dissipation. 
Once these structures are formed, their motion affects the dynamics of the 
surrounding species, thereby realizing a set of co-evolving boundary conditions.
This is how soft reactive matter implements the downward causal branch
invoked by the "weak" form of strong emergence. 
The driving engine behind this scenario is the peculiar ability of nonlinear
interactions to transfer mass, energy and information across scales,  from small to
large (coalescence) and back (breakup).  This multiscale coupling is the physical
mechanism implementing down-ward causality.

\subsection{Self-catalytic DPD}

Self-catalytic Advection-Diffusion-Reaction networks 
provide a rich  source of structure formation, hopefully beyond the level 
of Turing patterns and dissipative structures, hence they may 
reach farther towards the goal of explaining structure 
formation at the level of the organs.
I mean systems of DPD-like equations,  
augmented with self-catalytic reactions of the form:
$A+B \to C; \; C+D \to E;\; E+F \to A$.
Note that none of the three elementary steps above is 
auto-catalytic, but the set of the three together
is because at the end of the loop,  A returns to A. 
This kind of reactions are conducive to a key property of 
living systems, i.e.  the ability to replicate and eventually 
reproduce. The spatial interactions are then in charge of securing that
such replication/reproduction effects take place only where and 
when needed.
Much work has been done on auto-catalytic networks,  but I am not
aware of any merger with mesoscale reactive particle dynamics (itself a dynamic network).
Once mesoparticles are equipped with suitable interactions supporting surface tension
and other forms of dispersion forces, along with dissipation and suitable
self-catalytic reactions,  such combination might take us one step forward
towards the emergence of functional structures.

Formally,  they look as a series of reactive DPD's, one for each chemical species:
\begin{eqnarray}
\label{READPD}
\frac{dR_{I,s}}{dt} = V_{I,s}\\
\frac {d( M_{I,s} V_{I,s})}{dt} = \sum_{J>I,s'>s} 
F_{IJ}^{C,ss'} + F_{IJ}^{D,ss'} + F_{IJ}^{R,ss'}\\
\frac{dM_{I,s}}{dt} = \mathcal{R}_{I,s}\;\;\;I=1,N\;s=1,N_S
\end{eqnarray}
Can such reactive DPD's  ever reach the complexity of biological
organs, such as the heart or the eye,  i.e. {\it physiological complexity}?
I, for one, would be inclined towards a sharp no, but I think it is fair to concede
that until we figure out in quantitative detail what kind of structures 
can emerge from the above equations, the question remains open.
Interesting work along these lines is just beginning to appear \cite{DPDR}. 

\begin{figure}
\includegraphics[scale=0.5]{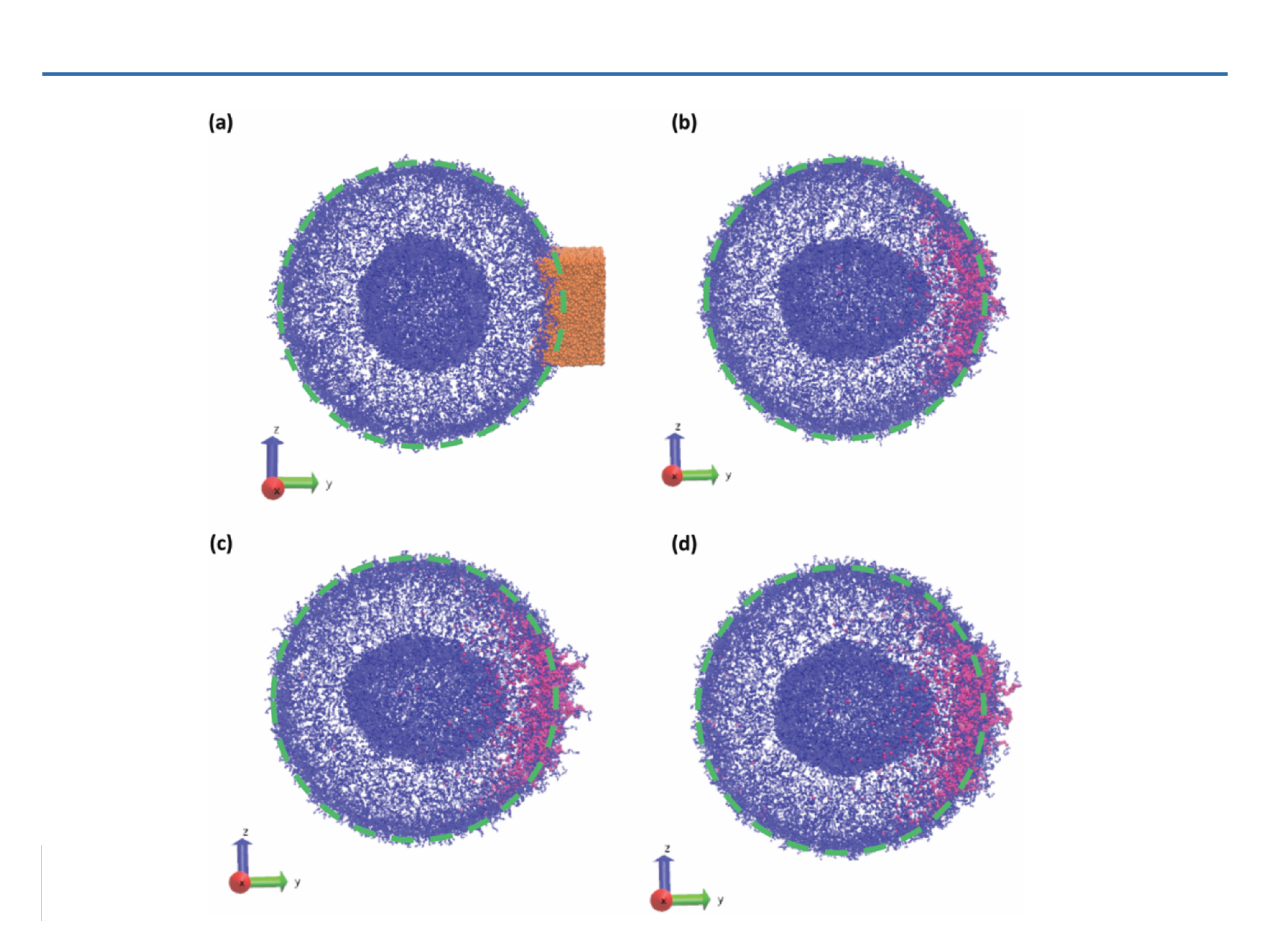}
\caption{Reactive DPD simulation of a shape-changing vesicle under the
effect of adsorption of solvophobic monomers (brown particles) 
at the outer region of the membrane. The chemical reactions at the outer membrane
convert the solvophobic monomers into solvophilic ones, with a consequent
hange in the shape of the vescicle.
From \cite{DPDR}.
}
\end{figure}

\section{Summary and Outlook}

In this Perspective I have tried to argue that modern statmech
has developed a number of remarkable ideas which shed new light
into mechanisms at the roots of biological complexity.
They can account for replication, non-ergodicity,  memory, 
non-locality, self-catalysis, co-evolving boundary 
conditions, and other fundamental mechanism typical of living
systems, not explicitly apparent from Newtonian physics. 

Is it enough to predict the emergence of the heart?
Is there room for "functional purpose" in our equations and algorithms?
Steven Weinberg famously sentenced the Universe as pointless:
"the more the Universe becomes comprehensible,  the more it 
appears pointless" \cite{WEIN}.  No room for purpose in the world of Lagrangians.
To which another towering figure, Freeman Dyson, retorted that
"no Universe with intelligence can be pointless" \cite{DYS}.
I stand by Dyson, on a blind date:
there is more to physics than Lagrangians.
The emergence of the cell first and of the
brain later, mark two subsequent boosts of complexity which
have changed the course of the natural world in ways that 
may indeed stand beyond reach of computable pre-stated equations,rules and algorithms.  
This is a fascinating and welcome speculation in many respects 
(who likes to be computable?).
Yet,  until  we know yet what levels of complexity can
be effectively attained by the systems I have tried to describe 
in this Perspective, this question remains up for grabs.

\section{Personal disclaimer}

Whenever Science investigates the functioning of living systems, it
gets inevitably in close touch with Religion, a close
encounter which for many spells inevitable conflict.
I am not among these: in my view, Science and Religion 
address different instances of human quest and they should proceed
independently, with great mutual respect.
The fact that physics may one day explain evolutionary
properties of matter, including the emergence of complex functional units
like organs, does by no means rule out the existence of a Superior Being.
Quite on the contrary, it would point to a plan that comes 
with both hardware and software, which only makes it more 
subtle, elegant and worth admiration and gratitude.
Georges Lemaitre says it best \cite{LEM}:
{\it
"the whole story of the world need not have been written 
down in the first quantum like a song on the disc of a phonograph. 
The whole matter of the world must have been present at the 
beginning, but the story it has to tell may be written step by step"}.

\section{Acks}
The author kindly acknowledges illuminating discussions with 
Stuart Kauffman, from which this article originated.
He also wishes to thank M. Cortes, P. Coveney, G. Ellis, D. Frenkel,A. Irback, A. Liddle G. Parisi and P. Tello
for very useful exchanges.  Funding from the European Research
Council under the Horizon 2020 Programme Grant Agreement n. 739964 ("COPMAT")
is gratefully acknowledged.


\begin{thebibliography}{99}



\bibitem{EPS} Physics of living matter,
Special issue of europhysics news, EPN 51/5 (2020)

\bibitem{KAUF1} S. Kauffman, 
A World Beyond Physics, Oxford U.P. (2019)

\bibitem{KAROL} SA  Kauffman,  A. Roli,
The world is not a theorem, aXiv preprint arXiv:2101.00284 (2021)

\bibitem{PWA} P.W. Anderson,
More is Different,
Science, New Series, Vol. 177, No. 4047., 393-396 (1972).

\bibitem{PCB} JP Boon, P. Coveney and S. Succi,
Multiscale models at the physics-chemistry-biology interface,
Phil. Trans. Roy. Soc., 374,20160335, (2016) 

\bibitem{GIB}  J. J. Gibson, 
The Ecological Approach to Visual Perception. 
Houghton Mifflin Harcourt (HMH), Boston (1979) 

\bibitem{KOHN} W. Kohn,
Nobel Lecture: Electronic structure of matter - wave functions and density functionals, 
Reviews of Modern Physics, Vol. 71,  1253  (1999)

\bibitem{WOLF} S. Wolfram,
A project to find the fundamental theory of physics,
Wolfram Media Inc, (2020)

\bibitem{SPIN} M. Mezard,  G. Parisi, M. Virasoro, 
Spin glass theory and beyond, Singapore, World Scientific, 1987, ISBN 9971-5-0115-5.
Phys. Rev. A 38, 364 - (1987)
 

\bibitem{THU} S. Thurner, R. Hanel and P. Klimek,
An introduction to the theory of complex systems,
Oxford Univ. Press,  (2018)
 
\bibitem{YOLO} Y.  Holovatc,  R. Kenna and S. Thurner,
Complex systems: physics beyond physics,
European Journal of Physics, Volume 38, Number 2 (2017)

\bibitem{GROS} C Gros,
Complex and Adaptive Dynamical Systems, Springer: Complexity, (2008)

\bibitem{TAP} I owe this clarification to Marina Cortes and Andrew Liddle.

\bibitem{ELLIS} G. Ellis,
Emergence in Solid State Physics and Biology,
Foundations of Physics, 50, 1098 (2020)

\bibitem{THERMO} B. Smit and D. Frenkel,
Understanding Molecular Simulation: from Algorithms to Applications,
Elsevier,  New York, (2001)

\bibitem{DPD1} P. J. Hoogerbrugge and J. M. V. A. Koelman,
Simulating Microscopic Hydrodynamic Phenomena with Dissipative Particle Dynamics
EPL (Europhysics Letters), Volume 19, Number 3 (1992)

\bibitem{DPD2} RD Groot, PB Warren, 
Dissipative particle dynamics: Bridging the gap between atomistic and mesoscopic simulation,
J. Chem. Phys. 107, 4423 (1997)

 \bibitem{SS22} S. Succi 
Sailing the Ocean of Complexity: Lessons from the Physics-Biology Frontier, 
Oxford U.P.,  in press 2022 

\bibitem{WOLYN} JD Bryngelson, JN Onuchic, ND Socci, PG Wolynes,
Funnels, pathways, and the energy landscape of protein folding: a synthesis
Proteins: Structure, Function, and Bioinformatics 21 (3), 167-195	(1995)

\bibitem{SINAI} Y. Sinai,
Dynamical systems with elastic reflections. Ergodic properties of dispersing billiards,
Russ. Math. Surv. 25:137-189 (1970).

\bibitem{GALLA} G. Gallavotti,
Ergodicity, ensembles, irreversibility in Boltzmann and beyond,
J.  of Stat. Phys. 78(5), 1571, (1995)

\bibitem{BOG} BM Boghosian, PV Coveney, H  Wang, 
A new pathology in the simulation of chaotic dynamical systems on digital computers,
Adv. Theory Simul. 2, 1900125. (doi:10.1002/ adts.201900125) (2019)

\bibitem{TUR} A. Turing, 
The Chemical Basis of Morphogenesis,
Philosophical Transactions of the Royal Society of London B. 237 (641): 37-72 (1952)
 
\bibitem{PRIGO1} I. Prigogine,
From Being to Becoming: Time and Complexity in the Physical Sciences,
Freeman and Company, (1980)

\bibitem{PRIGO2} I. Prigogine,
Modern Thermodynamics: From Heat Engines to Dissipative Structures,
Wiley and Sons, (2014)

\bibitem{WHAT} E. Schroedinger,
What Is Life? : The Physical Aspect of the Living Cell, Cambridge Univ. Press, (1944)

\bibitem{CROOK} GE Crooks
Entropy production fluctuation theorem and the nonequilibrium 
work relation for free energy differences,
Physical Review E 60 (3), 2721, (1999)

\bibitem{JAR} C Jarzynski
Nonequilibrium equality for free energy differences
Physical Review Letters 78 (14), 2690	4924	(1997)

\bibitem{GALCO} G Gallavotti, EGD Cohen,
Dynamical ensembles in nonequilibrium statistical mechanics,
Physical Review Letters 74 (14), 269 (1995)

\bibitem{JENG13} JL England,
Statistical physics of self-replication,
The Journal of Chemical Physics 139 (12), 09B623-1424	(2013)

\bibitem{JENG15} JL England,
Dissipative adaptation in driven self-assembly,
Nature Nanotechnology 10, 919-923,	(2015)

\bibitem{LB18}  S. Succi, 
The Lattice Boltzmann Equation for Complex States of Flowing Matter,
Oxford U.P.,  (2018) 

\bibitem{AM19}  A Montessori, M Lauricella, N Tirelli, S Succi,
Mesoscale modelling of near-contact interactions for complex flowing interfaces
Journal of Fluid Mechanics, 872, 327-347	30,	(2019)

\bibitem{DPDR} Qinyu Zhu, Timothy R. Scotta  and  Douglas R. Tree, 
Using reactive dissipative particle dynamics to understand local shape 
manipulation of polymer vesicles,  Soft Matter,  17, 24-39, (2021)
 
\bibitem{WEIN} S. Weinberg, 
The First Three Minutes: A Modern View of the Origin of the Universe, (1983)
ISBN 0465024378.

\bibitem{DYS} F.  Dyson,
A Many-Colored Glass: Reflections on the Place of Life in the Universe,  (2010)

\bibitem{LEM} G. Lemaitre,
The Beginning of the World from the Point of View of Quantum Theory,
Nature, 3210(127), 447 (1931)







 
%




 


 





\end{thebibliography}
\end{document}